\documentclass[superscriptaddress,showpacs,preprint]{revtex4}
\usepackage{amssymb}
\usepackage[tbtags]{amsmath}
\usepackage{graphicx}
\usepackage{epsfig,graphicx,times}

\begin{document}

\title{Gains Without Inversion in Quantum Systems with Broken Parities}
\author{W. Z. Jia}
\email{jiawenz1979@126.com}
\affiliation{Quantum Optoelectronics Laboratory, Southwest Jiaotong University, Chengdu
610031, China\\
}
\author{L. F. Wei}
\email{lfwei@home.swjtu.edu.cn}
\affiliation{Quantum Optoelectronics Laboratory, Southwest Jiaotong University, Chengdu
610031, China\\
}
\date{\today}

\begin{abstract}
For a quantum system with broken parity symmetry, selection rules can not
hold and cyclic transition structures are generated. With these
loop-transitions we discuss how to achieve inversionless gain of the probe
field by properly setting the control and auxiliary fields. Possible
implementations of our generic proposal with specific physical objects with
broken parities, e.g., superconducting circuits and chiral molecules, are
also discussed.
\end{abstract}

\maketitle

\bigskip




\bigskip

\section{Introduction}

It is well known that under the usual electric-dipole approximation, natural
atoms obey the optical selection rules, since their quantum states have
well-defined parity symmetries. With these electric-dipole transitions,
strong interactions between the fields and atoms have been utilized to
dynamically manipulate quantum coherence. As a consequence, many interesting
optical phenomena, such as coherent population trapping (CPT) \cite{CPT},
electromagnetically induced transparency (EIT)~\cite{EIT}, lasing without
inversion (LWI)~\cite{LWI} and so on, can be implemented. Basically, these
phenomena are originated from the absorption cancelation via quantum
interference of various allowed dipole transitions. Specifically, LWI
provides an approach to demonstrate optical gain without requiring
population inversion of atomic levels.

Besides the usual electric-dipole transitions, various relatively-weak
magnetic-dipole transitions are also utilized to realize certain
transitions~forbidden by electric-dipole selection rules. Typically, for a
natural three-level atom two electric-dipole transitions (between quantum
states with different parities, e.g., $|1\rangle $ and $|2\rangle $, and $%
|2\rangle $ and $|3\rangle $) and a magnetic-dipole transition (between near
degenerate quantum states with same parities, i.e. $|1\rangle $ and $%
|3\rangle $) can generate a loop-transition structure \cite%
{LOOPinATOM-CPT,LOOPinATOM-EIT,LOOPinATOM-GVC,LOOPinATOM-LWI}. Such
loop-transition configurations have been used to control phenomena
associated with\ atomic coherence, including CPT \cite{LOOPinATOM-CPT}, EIT
\cite{LOOPinATOM-EIT}, group velocity control \cite{LOOPinATOM-GVC} and LWI
\cite{LOOPinATOM-LWI}.

Recently, certain quantum systems with broken parity symmetries had been
investigated. These systems include, e.g., chiral molecules~\cite%
{Shapiroprl87,Shapiroprl90,LYprl}, asymmetric quantum wells~\cite%
{Shapiroprl87}, and superconducting quantum circuits (SQCs)~\cite%
{Clarke88,LYXprl,exp1,exp2}, etc. Both quantum bound states and interaction
Hamiltonian (transition matrix elements) in these parity-broken systems have
not well defined parities, and thus the usual selection rules do not hold.
Therefore, certain particular transition structures, e.g., three-level $%
\Delta $-type cyclic transition, can be realized \cite%
{Shapiroprl87,Shapiroprl90,LYprl,LYXprl}. Compared with loop structure in
natural atoms, in parity-broken systems, even three levels are\ well
separated from each other, the possible transition channels can form a loop
for selection rules do not hold. Note that such a configuration has been
experimentally demonstrated with circuit quantum electrodynamics (QED)
systems~\cite{exp1}, and has already been utilized to achieve tunable
coupling between two flux qubits~\cite{exp2}.

\begin{figure}[b]
\includegraphics[width=0.25\textwidth]{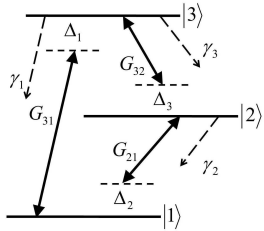}
\caption{Three-level quantum system with broken parity. The triangle-shaped
transition structure allows three possible transition channels coherently
driven by the fields with Rabi frequencies $G_{ij}$ and detunings $\Delta
_{i},\,i,j=1,2,3$. $\protect\gamma _{i}$ are the decay rates of the
corresponding levels.}
\end{figure}

In this paper, we investigate how to generate gain without inversion in
parity-broken three-level quantum systems by utilizing the $\Delta $-type
transition structure. We consider a $\Delta $-type three-level system
interacting simultaneously with three external fields; a weak probe, a
strong coherent control, and a tunable auxiliary ones. The optical response
of a quantum system with broken-parity symmetry is sensitive to the relative
phase of the three coherent driving fields. We show that the desirable
inversionless gains (called lasers or masers) of the weak probe field can be
achieved by properly controlling the parameters of applied driving fields.

Compared with the previous schemes for realizing LWIs with the
loop-transition configurations in \textit{microscopic} natural atoms~\cite%
{LOOPinATOM-LWI}, we emphasize that: (i)the LWI could also be demonstrated
with certain macroscopic quantum systems, such as SQCs; (ii) gain without
inversion can be generalized from the traditionally optical waveband (with
natural atoms) to the microwave domain (since the energy splittings of SQCs
are just in this waveband); and (iii) manipulating the cyclic transitions in
the present parity-broken artificial atoms to realize the LWI is relatively
simple. This is because one of three transitions in loop configuration with
natural atoms is usually implemented by using a significantly-weak
magnetic-dipole transition. However, in the loop structure with
parity-broken artificial atoms, the strengths of three transitions could be
at the same orders. 

The paper is organized as: In Sec. II we firstly give an universal analysis
on the gain-absorption properties in quantum systems with parity-broken
symmetries. Then, in Sec. III, we discuss how to demonstrate our generic
proposals with two class specific physical systems, the superconducting flux
qubits and the chiral molecules. Conclusions and discussions are given in
Sec. IV.

\section{Gain without population inversion in quantum systems\ with $\
\Delta $-type cyclic transitions}

\begin{figure}[b]
\includegraphics[width=0.5\textwidth]{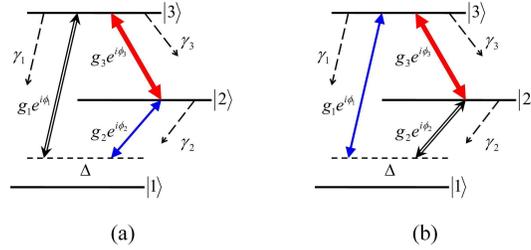}
\caption{{(Color online) A $\Delta $-type\ three-level system driven by
three coherent fields: (a) A strong coupling field (red line) applied
resonantly to the transition channel $\left\vert 2\right\rangle
\longleftrightarrow \left\vert 3\right\rangle $, a weak probe (blue line)
and another auxiliary field ($\Leftrightarrow $ line) with equal detunings
are applied to the\ transition channels $\left\vert 1\right\rangle
\longleftrightarrow \left\vert 2\right\rangle $ and $\left\vert
1\right\rangle \longleftrightarrow \left\vert 3\right\rangle $,
respectively. (b) A strong coupling field applied resonantly to the
transition channel $\left\vert 1\right\rangle \longleftrightarrow \left\vert
2\right\rangle $, a weak probe and another auxiliary field with equal
detunings are applied to the transition channels $\left\vert 2\right\rangle
\longleftrightarrow \left\vert 3\right\rangle $ and $\left\vert
1\right\rangle \longleftrightarrow \left\vert 3\right\rangle $, respectively.%
}}
\label{fig2}
\end{figure}
Consider a three-level system with a cyclic transition structure shown in
Fig.~1. In our inversionless gain scheme with parity-broken three-level
systems, three coherent driving fields are applied; one is the strong
resonant-coupling field, one is applied as a probe, and the third one is a
tunable auxiliary field. Depending on the specific systems adopted, these
three applied fields could be either microwaves or optical waves. The
Hamiltonian of our generic system can be written as $H=\sum_{i=1}^{3}E_{i}%
\left\vert i\right\rangle \left\langle i\right\vert +\frac{1}{2}%
\sum_{i>j=1}^{3}\left[ G_{ij}e^{i\omega _{ij}t}\left\vert i\right\rangle
\left\langle j\right\vert +\mathrm{H.c.}\right] $, where $E_{i}$ are the
eigenvalues of energy eigenstates $\left\vert i\right\rangle $, $\omega
_{ij} $ and $G_{ij}$ are the frequencies and Rabi frequencies of the
coherent driving fields, respectively. Let $\Delta _{1}=E_{3}-E_{1}-\omega
_{31}$, $\Delta _{2}=E_{2}-E_{1}-\omega _{21}$ and $\Delta
_{3}=E_{3}-E_{2}-\omega _{32}$ be the detunings of the applied driving
fields. When the condition $\Delta _{1}=\Delta _{2}+\Delta _{3}$ is
satisfied and in interaction picture, the interaction Hamiltonian is time
independent: $H_{I}=\Delta _{1}\left\vert 3\right\rangle \left\langle
3\right\vert +\Delta _{2}\left\vert 2\right\rangle \left\langle 2\right\vert
+\frac{1}{2}\sum_{i>j=1}^{3}\left[ G_{ij}\left\vert i\right\rangle
\left\langle j\right\vert +\mathrm{H.c.}\right] $. The dynamical evolution
of the system, including relaxation terms $\gamma _{i}$, is governed by the
Liouvillian equation: $d\rho /dt=-i\left[ H_{I},\rho \right] +\mathcal{L}%
\left[ \rho \right] $.

The closed-loop atomic configuration considered here makes the
optical properties of system are sensitive to the relative phases of
applied fields \cite{LOOPinATOM-EIT}. The Rabi frequencies $G_{ij}$
should be dealt as complex parameters, e.g., $G_{31}/2=g_{1}e^{i\phi
_{1}}$, $G_{21}/2=g_{2}e^{i\phi_{2}}$, and
$G_{32}/2=g_{3}e^{i\phi_{3}}$, with $2g_{i} $ being the amplitudes
and $\phi _{i}$ the phases. Redefining the density matrix elements:
$\rho _{ii}=\sigma _{ii}$, $\rho _{13}=\sigma _{13}e^{-i\phi _{1}}$,
$\rho _{23}=\sigma _{23}e^{-i\phi _{3}}$, $\rho _{12}=\sigma
_{12}e^{i\left( \phi _{3}-\phi _{1}\right) }$, and from the above
Liouvillian equation, one can obtain their equations of motion:
\begin{eqnarray}
\overset{.}{\sigma }_{11} &=&\gamma _{1}\sigma _{33}+\gamma _{2}\sigma
_{22}-(ig_{1}\sigma _{13}+ig_{3}\sigma _{12}e^{i\Phi }+\mathrm{H.c.}), \\
\overset{.}{\sigma }_{22} &=&-\gamma _{2}\sigma _{22}+\gamma _{3}\sigma
_{33}+(ig_{2}\sigma _{12}e^{i\Phi }-ig_{3}\sigma _{23}+\mathrm{H.c.}), \\
\overset{.}{\sigma }_{12} &=&\left( -\Gamma _{12}+i\Delta _{2}\right) \sigma
_{12}-ig_{3}\sigma _{13}+ig_{1}\sigma _{32}+ig_{2}e^{-i\Phi }\left( \sigma
_{22}-\sigma _{11}\right) , \\
\overset{.}{\sigma }_{13} &=&\left( -\Gamma _{13}+i\Delta _{1}\right) \sigma
_{13}+ig_{1}\left( \sigma _{33}-\sigma _{11}\right) +ig_{2}\sigma
_{23}e^{-i\Phi }-ig_{3}\sigma _{12}, \\
\overset{.}{\sigma }_{23} &=&\left( -\Gamma _{23}+i\Delta _{3}\right) \sigma
_{23}+ig_{2}\sigma _{13}e^{i\Phi }-ig_{1}\sigma _{21}+ig_{3}\left( \sigma
_{33}-\sigma _{22}\right) ,
\end{eqnarray}%
where $\Phi =\left( \phi _{2}+\phi _{3}-\phi _{1}\right) $ is the relative
phase of the applied fields, and $\Gamma _{12}=\gamma _{2}/2$, $\Gamma
_{13}=\left( \gamma _{1}+\gamma _{3}\right) /2$, $\Gamma _{23}=\left( \gamma
_{1}+\gamma _{2}+\gamma _{3}\right) /2.$

For simplicity, we assume that all the decay rates of the levels are equal,
namely $\gamma _{i}=\gamma $. 
The steady-state solution of the master equation can be attained by setting $%
\overset{.}{\sigma}_{ij}=0$. 
We first consider the case shown in Fig.~2(a). A resonant coupling field $%
g_{3}$ is applied to the transition between the intermediate state $%
\left\vert 2\right\rangle $ and the upper state $\left\vert 3\right\rangle $%
, a weak coherent field $g_{2}$ with the detuning $\Delta $ acted as a probe
is applied to the transition between the ground state $\left\vert
1\right\rangle $ and intermediate state $\left\vert 2\right\rangle $, and an
auxiliary field $g_{1}$ with the same detuning $\Delta $ couples the levels $%
\left\vert 1\right\rangle $ and $\left\vert 3\right\rangle $, respectively.
\begin{figure*}[t]
\includegraphics[width=0.8\textwidth]{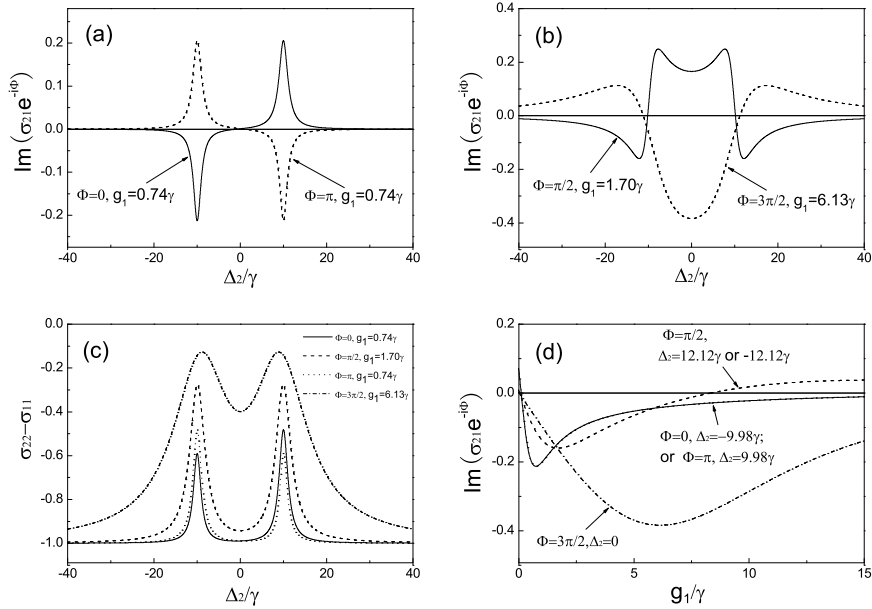}
\caption{Phase sensitive probe gain without population inversion of the
configuration shown in Fig.~2(a), with the parameters $\protect\gamma _{i}=%
\protect\gamma $, $g_{3}=10\protect\gamma $ and $g_{2}=0.1\protect\gamma $:
(a) Detuning-dependent probe gain for $g_{1}=0.74\protect\gamma $ and $\Phi
=0$ (solid line); $\protect\pi $ (dashed line). (b) Detuning-dependent probe
gain for $\Phi =\protect\pi /2,\,g_{1}=1.70\protect\gamma $ (solid line) and
$\Phi =3\protect\pi /2,\,g_{1}=6.13\protect\gamma $ (dashed line). (c)
Detuning-dependent population difference $\protect\sigma _{22}-\protect%
\sigma _{11}$ corresponds to the probe gains shown in (a) and (b). It is
clearly shown that these gains are not due to the population inversions. (d)
Probe gain versus amplitude of Rabi frequency of auxiliary field $g_{1}$.
With fixed relative phases and detunings, maximal gains can be gotten at $%
g_{1}=0.74\protect\gamma $ (solid line);\ $1.70\protect\gamma $ (dashed
line); or $6.13\protect\gamma $ (dash-dotted line). This means that the
parameters of auxiliary field utilized in (a) and (b) for getting the gains
are optimal.}
\label{fig3}
\end{figure*}
The absorption behavior for the probe $g_{2}$ can be described by $\mathrm{Im%
}(\sigma _{21}e^{-i\Phi })$. For the configuration displayed in Fig.~2(a),
it is seen, from Figs.~3(a) and (b), that remarkable gains, i.e., $\mathrm{Im%
}(\sigma _{21}e^{-i\Phi })<0$, can be established, if the modulus of Rabi
frequency $g_{1}$ and the relative phase $\Phi $ are modulated
appropriately. In fact, the modulation of $\Phi $ can be achieved by fixing $%
\phi _{2}$, $\phi _{3}$ and changing the phase of the auxiliary field $\phi
_{1}$ only. Typically, when $g_{1}=0.74\gamma $ and $\Phi =0$ (or $\pi )$,
it is shown clearly that gain dip appears at $\Delta _{2}\approx -9.98\gamma
$ (or $9.98\gamma $); when $\Phi =\pi /2,\,g_{1}=1.70\gamma $, two gain
regions locate respectively at about $\Delta _{2}<-10\gamma $ and $\Delta
_{2}>10\gamma $, with the maximum gain points appearing at $\Delta
_{2}\approx \pm 12.12\gamma $; and when $\Phi =3\pi /2$, $g_{1}=6.13\gamma $%
, a remarkable probe gain can be established approximately in a wide
spectral range from about $-10\gamma $ to $10\gamma $, with the maximum gain
point being located at $\Delta _{2}=0$. Fig.~3 (d) shows that the Rabi
frequencies of the auxiliary fields, used in Figs.~3(a) and 3(b), are
optimal for implementing the desirably maximum gains. More interestingly,
Fig.~3(c) shows that population inversion $\sigma _{22}-\sigma _{11}$ is
always less than zero for any detuning. This indicates that the
phase-dependant gains attained in Fig.~3(a) and (b) are inversionless.
\begin{figure*}[t]
\includegraphics[width=0.8\textwidth]{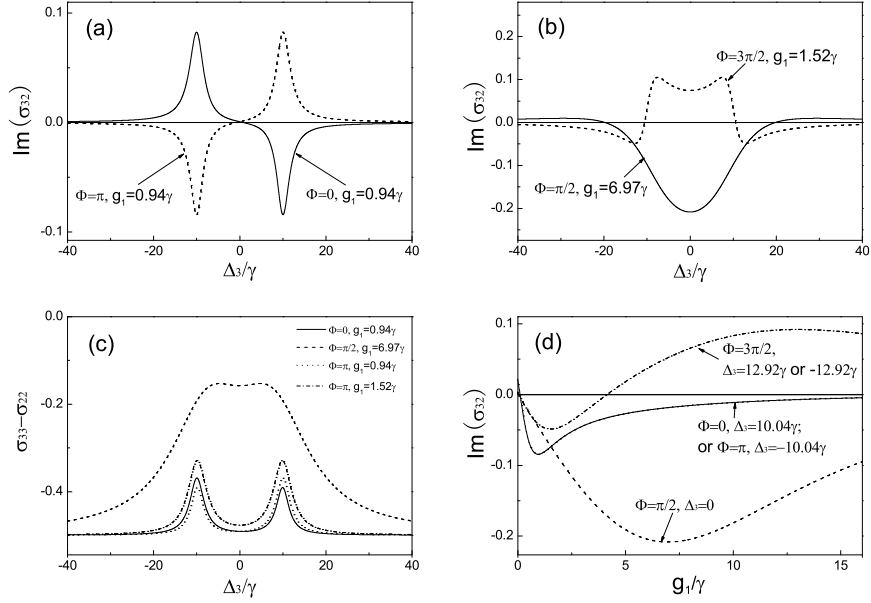}
\caption{Phase sensitive probe gain without population inversion of the
configuration shown in Fig.~2(b), with the parameters $\protect\gamma _{i}=%
\protect\gamma $, $g_{2}=10\protect\gamma $ and $g_{3}=0.1\protect\gamma $:
(a) Detuning-dependent probe gain for $g_{1}=0.94\protect\gamma $ and $\Phi
=0$(solid line); $\protect\pi $(dashed line). (b) Detuning-dependent probe
gain for $\Phi =\protect\pi /2,\,g_{1}=6.97\protect\gamma $ (solid line) and
$\Phi =3\protect\pi /2,\,g_{1}=1.52\protect\gamma $ (dashed line). (c)
Detuning-dependent population difference $\protect\sigma _{33}-\protect%
\sigma _{22}$ corresponds to the probe gains shown in (a) and (b). It is
clearly shown that these gains are not due to the population inversions. (d)
Probe gain versus amplitude of Rabi frequency of auxiliary field $g_{1}$.
With fixed relative phases and detunings, maximal gains can be gotten at $%
g_{1}=0.94\protect\gamma $ (solid line);\ $6.97\protect\gamma $ (dashed
line); or $1.52\protect\gamma $ (dash-dotted line). This means that the
parameters of auxiliary field utilized in (a) and (b) for getting the gains
are optimal.}
\label{fig4}
\end{figure*}

On the other hand, if the atomic configuration in Fig.~2(b) is selected,
i.e., $g_{2}$ acts as a resonant coupling field,$\ g_{3}$ is a weak probe
with detuning $\Delta $, and $g_{1}$ is an auxiliary field with the same
detuning, Figs.~4(a-c) show similarly that the phase-sensitive gain without
population inversion could still be achieved. Specifically, when $%
g_{1}=0.94\gamma $ and $\Phi =0$ (or $\pi $), it is shown clearly that gain
dip appears at $\Delta _{3}\approx 10.04\gamma $ (or $-10.04\gamma $); when $%
\Phi =\pi /2,\,g_{1}=6.97\gamma $, a remarkable probe gain can be
established in a spectral range from about $-20\gamma $ to $20\gamma $, with
the maximum gain point being located at $\Delta _{3}=0$; and when $\Phi
=3\pi /2$, $g_{1}=1.52\gamma $, two gain regions locate respectively at
about $\Delta _{3}<-10\gamma $ and $\Delta _{3}>10\gamma $, with the maximum
gain points appearing at $\Delta _{3}\approx \pm 12.92\gamma $. Also, for
the fixed $\Phi $ and detuning $\Delta _{3}$, Fig.~4(d) shows how the $%
\mathrm{Im}(\sigma _{32})$ depends on the parameter $g_{1}$. One can see
that the parameters selected in Figs.~4(a-c) are optimal for realizing the
gains.

The above numerical results can be simply explained by investigating the
steady-state condition in Eqs.~(3) and (5):
\begin{eqnarray}
&&\mathrm{Im}\left( \sigma _{21}e^{-i\Phi }\right) =g_{2}\Gamma _{12}\left(
\sigma _{11}-\sigma _{22}\right) /\mathcal{A}+\mathrm{Im}\left[ \frac{1}{%
\mathcal{A}}\left( \Gamma _{12}-i\Delta _{2}\right) \left( ig_{3}\sigma
_{31}-ig_{1}\sigma _{23}\right) e^{-i\Phi }\right] , \\
&&\mathrm{Im}\sigma _{32}=g_{3}\Gamma _{23}\left( \sigma _{22}-\sigma
_{33}\right) /\mathcal{B}+\mathrm{Im}[\left( \Gamma _{23}-i\Delta
_{3}\right) \left( ig_{1}\sigma _{12}-ig_{2}\sigma _{31}e^{-i\Phi }\right) /%
\mathcal{B}]
\end{eqnarray}%
with $\mathcal{A}=\Gamma _{12}^{2}+\Delta _{2}^{2}$, $\mathcal{B}=\Gamma
_{23}^{2}+\Delta _{3}^{2}$. Clearly, in the configuration shown in Fig.
2(a), in order to achieve gain without inversion for probe $g_{2}$ the
conditions $\mathrm{Im}\left( \sigma _{21}e^{-i\Phi }\right) <$0, and $%
\sigma _{11}-\sigma _{22}>0$ should be simultaneously satisfied. It can be
seen from Eq. (6) that, when $\sigma _{11}-\sigma _{22}>0$, the first term
on the right-hand side of Eq.~(6) is positive. Thus, in order to get a gain,
namely $\mathrm{Im}\left( \sigma _{21}e^{-i\Phi }\right) <0$, the second
term on the right-hand side of Eq.~(6) must contribute negatively to $%
\mathrm{Im}\left( \sigma _{21}e^{-i\Phi }\right) $. This implies that in
this case the induced inversionless gain is originated from the dynamically
induced coherence by the coupling field $g_{3}$ and the auxiliary one $g_{1}$%
. Clearly, if the auxiliary field is not applied, the system shown in Fig.
2(a)\ is reduced to the usual ladder-type configuration. In the presence of
a control field\ $g_{3}$ and a probe field $g_{2}$, the phenomenon of EIT
can be achieved instead of inversionless gain. However, in the present
systems with $\Delta $-type cyclic transition structure (due to the broken
parity symmetries), an auxiliary coherent driving field $g_{1}$ could be
applied to couple the levels $\left\vert 1\right\rangle $ and $\left\vert
3\right\rangle $. As a consequence, the term related to $\sigma _{23}$
appears and thus negative values of $\mathrm{Im}\left( \sigma _{21}e^{-i\Phi
}\right) $ can be induced within certain spectral ranges. This is clearly
proven that the auxiliary driving field plays crucial roles for the
appearance of the gain. Similarly, Eq. (7) shows that for the configuration
displayed in Fig.~2(b), an auxiliary field $g_{1}$ is necessary to obtain
the phase-dependant gain without inversion of the probe field $g_{3}$.

\section{Physical demonstrations and possible applications}

The above generic results can be realized with all the systems whose quantum
states possess broken parity symmetries, such as chiral molecules~\cite%
{Shapiroprl87,Shapiroprl90,LYprl}, asymmetric quantum wells~\cite%
{Shapiroprl87}, superconducting quantum circuits (SQCs)~\cite%
{Clarke88,LYXprl,exp1,exp2}, and so on.

Typically, SQCs can be regarded as artificial atoms with quantized energy
levels. Quantum-mechanical behaviors in these artificial atoms, such as
spectroscopy~\cite{spe}, Rabi oscillations~\cite{Rabi}, and so forth, have
already been demonstrated experimentally. Also, SQCs coupling to various
bosonic modes (e.g., microwave fields, nano-mechanical resonator \cite{NHR},
superconducting\ transmission line \cite{TL}, etc.) can be utilized to
simulate various quantum optical phenomena in the microwave domain.
Moreover, recent studies show that quantum optical phenomena related to
atomic coherence, such as EIT~\cite{EITinSQC1,EITinSQC2}, Autler-Townes
effects~\cite{ATinSQC} and CPT \cite{CPTinSQC}, can also be achieved in
SQCs. Here, by using their special loop transition-structure we show that
these devices could be utilized to realize another important phenomenon
related to atomic coherence, i.e., the microwave gain without population
inversion mentioned above.

In SQCs the desirable cyclic transition configurations could be demonstrated
with both flux and phase qubits~\cite{Clarke88,LYXprl,exp1,exp2}. In fact,
the parity-broken and consequently loop-transition configurations had been
first observed in phase-qubit experiments~\cite{Clarke88} and recently
demonstrated with flux qubits~\cite{exp1,exp2}. Here, we take the flux
qubits as typical examples. 
For the artificial atoms generated by three Josephson-junction
circuits selection rules do not always exist, as the parity of the
system can be broken by adjusting the parameters of circuits. The
effective potential of the system reads (See, e.g.,
Refs.~\cite{LYXprl,EITinSQC1}.) $U(\varphi _{m},\varphi
_{p})=2E_{J}\left( 1-\cos \varphi _{p}\cos \varphi _{m}\right)
+\alpha E_{J}\left[ 1-\cos \left( 2\pi f+2\varphi _{m}\right)
\right] $, with $\varphi _{p,m}=\left( \varphi _{1}\pm \varphi
_{2}\right) /2$ being
the generalized coordinates defined by the phase drops $\varphi _{1}$ and $%
\varphi _{2}$ across the two larger junctions, respectively, and $E_{J}$
their Josephson energies. The reduced magnetic flux $f=\Phi _{e}/\Phi _{0}$
is defined as the ratio of the external magnetic flux $\Phi _{e}$ with the
flux quantum $\Phi _{0}$, and $0<\alpha<1$. Clearly, if the flux is biased
away from the degenerate point with $f=1/2$, the potential $U(\varphi
_{m},\varphi _{p})$ has ill-defined parities and thus microwave-induced
transitions between arbitrary two levels are possible. 
Typically, it is seen from Ref.~\cite{LYXprl} that, for $f=0.496$ the lowest
three energy levels of the artificial atom are well separated from other
higher energy levels and the moduli $\left\vert t_{ij}\right\vert$ of
transition matrix elements between any two levers are comparable (i.e., $%
\left\vert t_{01}\right\vert \simeq 0.19$, $\left\vert t_{02}\right\vert
\simeq 0.14$, and $\left\vert t_{12}\right\vert \simeq 0.19$). This
indicates that the flux qubit is really an ideal candidate to realize the
phase-sensitive inversionless gain proposed in Sec. II.

The time scale to reach the steady-state solutions (see, Sec. II) required
for realizing LWI with the present SQCs could be estimated as follows. With
the experimentally-demonstrated relaxation rate $6.9\times 10^{7}s^{-1}$
(between the two lower levels of a flux qubit at degenerate point)~\cite%
{EITinSQC2,TsaiScience2010}, and the calculated transition matrix elements $%
\left\vert t_{ij}\left( f\right) \right\vert$ \cite{LYXprl}, the three decay
rates used in our proposal are estimated as: $\gamma _{2}=\gamma
_{3}=5.5\times 10^{6}\mathrm{s}^{-1}$, and $\gamma _{3}=3.2\times 10^{6}%
\mathrm{s}^{-1}$ (Note that the relaxation times are proportional to $%
\left\vert t_{ij}\right\vert ^{2}$). If the Rabi frequencies of coupling and
probe fields are setting as $10\gamma _{2}$ and $0.1\gamma_{2}$, and by
appropriately choosing Rabi frequency of auxiliary field, the time scale to
reach the so-called stationary solutions is numerically as $~10^{-6}$s. This
implies that the physical demonstration of our proposal with SQCs should be
feasible.

Furthermore, constructing an effective medium to realize inversionless maser
with these artificial atoms is also possible. In fact, the typical size of a
superconducting qubit is about 10$^{-6}$m \cite{Rabi}. This is much smaller
than the wavelength of microwave. For example, if a transition can be driven
by microwave field with wavelength around 10$^{-2}$m, the ratio between the
wavelength of driving field and the size of artificial atoms is about 10$%
^{4} $. This is in accordance with the ratio between the optical wavelength
and the size of natural atom. Therefore, an effective medium generated by a
block consisting of superconducting artificial atoms is feasible.

For another kind quantum system with broken parity symmetry, i.e., chiral
molecules \cite{Shapiroprl87,Shapiroprl90,LYprl}, our scheme proposed in
Sec. II, could also offer an effective way to discriminate the left- and
right-handed chiral molecules (Such pairs are called \textquotedblleft
enantiomers\textquotedblright\ \cite{Shapiroprl87}.), in addition to realize
LWI. If only the three lowest levels are considered, a chiral molecule can
be modeled as a three-level cyclic system as shown in Fig. 1. Thus three
lasers can be applied to enantiomeric molecules with the Rabi frequencies
being chosen as, for example, those in Fig. 3(a). The Rabi frequencies of
applied lasers between any pair of left- and right-handed states differ by a
sign, namely, $g_{i}e^{i\phi _{i}^{L}}=-g_{i}e^{i\phi _{i}^{R}}$ ($i=1,2,3$%
). Thus the according phase factors of Rabi frequencies $\phi _{i}^{L,R}$ ($%
i=1,2,3$) differ by $\pi $ \cite{Shapiroprl87}. Clearly, the difference
between the total phase factors of the two enantiomers is $\Phi ^{L}-\Phi
^{R}=\pi $, where $\Phi ^{L,R}=\phi _{2}^{L,R}+\phi _{3}^{L,R}-\phi
_{1}^{L,R}.$ On the other hand, as shown in Sec. II, the\ gain-absorption
properties of the two enantiomers (both with cyclic transition structures)
are dependent on the total phase factors $\Phi ^{L,R}.$ Clearly, if we set
the phase factor of the applied coherent fields appropriately to assure that$%
\ \Phi ^{L}=\pi $, then inevitably $\Phi ^{R}=0$. As a consequence, the
probe gain-absorption spectra of the\ left- and right-handed chiral
molecules correspond to the dashed and solid lines in Fig.~3(a),
respectively. Thus the enantiomers can be identified by their different
gain-absorption spectra.

\section{Conclusions and discussions}

In summary, we have shown that phase-sensitive gain without inversion can be
realized with parity-broken quantum systems. We investigate two typical
inversionless gain approaches by applying a probe, a coupling field, and a
tunable auxiliary field to generate a transition loop. In these approaches,
by modifying the phase and modulus of Rabi frequency of auxiliary field,
remarkable inversionless gains can be obtained for different probe
detunings. Our generic proposal could be implemented with various specific
systems with broken parity symmetries, e.g., superconducting artificial
atoms, chiral molecules, asymmetric quantum wells and so on. Therefore,
maser (laser) without inversion can be realized in principle with these
systems by using their cyclic transition structures. As far as chiral
molecules, the phase-dependent gain-absorption spectra may be used to
discriminate enantiomeric molecules.

\begin{acknowledgments}
The project was supported in part by National Natural Science
Foundation of China under Grant Nos. 10874142, 90921010, and the
National Fundamental Research Program of China through Grant No.
2010CB923104.
\end{acknowledgments}



\begin{thebibliography}{99}
\bibitem{CPT} E. Arimondo, in Progress in Optics XXXV, edited by E. Wolf
(North-Holland, Amsterdam, 1996).

\bibitem{EIT} S. E. Harris, J. E. Field, and A. Imamoglu, Phys. Rev. Lett.
\textbf{64} 1107 (1990); K. J. Boller, A. Imamoglu, and S. E. Harris, Phys.
Rev. Lett. \textbf{66}, 2593 (1991); S. E. Harris, Phys. Today \textbf{50,}
36 (1997); M. Fleischhauer, A. Imamoglu, and J. P. Marangos, Rev. Mod. Phys.
\textbf{77, }633 (2005).

\bibitem{LWI} S. E. Harris, Phys. Rev. Lett. \textbf{62} (1989) 1033; M. O.
Scully, S. Y. Zhu, and A. Gavrielides, Phys. Rev. Lett. \textbf{62}, 2813
(1989); A. Imamoglu, J. E. Field, S. E. Harris, Phys. Rev. Lett. \textbf{66}
1154 (1991); E. S. Fry, X. Li, D. Nikonov, G. G. Padmabandu, M. O. Scully,
A. V. Smith, F. K. Tittel, C. Wang, S. R. Wilkinson, and S. Y. Zhu, Phys.
Rev. Lett. \textbf{70}, 3235 (1993); A. S. Zibrov, M. D. Lukin, D. E.
Nikonov, L. Hollberg, M. O. Scully, V. L. Velichansky, and H. G. Robinson,
Phys. Rev. Lett. \textbf{75}, 1499 (1995); J. Mompart and R. Corbalan, J.
Opt. B: Quantum Semiclass. Opt. \textbf{2}, R7 (2000).

\bibitem{LOOPinATOM-CPT} D. V. Kosachiov, B. G. Matisov, and Yu. V.
Rozhdestvensky, Opt. Commun. \textbf{85}, 209 1991; D. V. Kosachiov, B. G.
Matisov and Yu. V. Rozhdestvensky, J. Phys. B \textbf{25}, 2473 (1992).

\bibitem{LOOPinATOM-EIT} B. P. Hou, S. J. Wang, W. L. Yu and W. L. Sun, J.
Phys. B, \textbf{38} 1419(2005); E. A. Wilson, N. B. Manson, C. Wei and L.
J. Yang, Phys. Rev. A \textbf{72}, 063813, (2005); Hebin Li, V. A.
Sautenkov, Y. V. Rostovtsev, G. R. Welch, P. R. Hemmer, and M. O. Scully,
Phys. Rev. A \textbf{80} 023820 (2009).

\bibitem{LOOPinATOM-GVC} D. Bortman-Arbiv, A. D. Wilson-Gordon, and H.
Friedmann, Phys. Rev. A \textbf{63} 043818 (2001).

\bibitem{LOOPinATOM-LWI} O. Kocharovskaya, P. Mandel, and Y. V. Radeonychev,
Phys. Rev. A \textbf{45} 1997 (1992); W. H. Xu, J. H. Wu\ and J. Y. Gao,
Opt. Commun. \textbf{223} 367 (2003).

\bibitem{Shapiroprl87} P. Kra'l and M. Shapiro, Phys. Rev. Lett. \textbf{87}%
, 183002 (2001).

\bibitem{Shapiroprl90} P. Kra'l, I. Thanopulos, M. Shapiro, and D. Cohen,
Phys. Rev. Lett. \textbf{90}, 033001 (2003).

\bibitem{LYprl} Yong Li, C. Bruder, and C. P. Sun, Phys. Rev. Lett. \textbf{%
99}, 130403 (2007).

\bibitem{Clarke88} J. M. Martinis, M. H. Devoret and J. Clarke, Phys. Rev. B
\textbf{35} 4682 (1987); J. Clarke, A. N. Cleland, M. H. Devoret, D. Esteve,
and J. M. Martinis, Science \textbf{239}, 992 (1988).

\bibitem{LYXprl} Yu-xi Liu, J. Q. You, L. F. Wei, C. P. Sun, and F. Nori,
Phys. Rev. Lett. \textbf{95}, 087001 (2005).

\bibitem{exp1} F. Deppe, M. Mariantoni, E. P. Menzel, A. Marx, S. Saito, K.
Kakuyanagi, H. Tanaka, T. Meno, K. Semba, H. Takayanagi, E. Solano, and R.
Gross, Nat. Phys. \textbf{4}, 686 (2008).

\bibitem{exp2} K. Harrabi, F. Yoshihara, A. O. Niskanen, Y. Nakamura, and J.
S. Tsai, Phys. Rev. B \textbf{79}, 020507(R) (2009).

\bibitem{spe} J. R. Friedman, V. Patel, W. Chen, S. K. Tolpygo, and J. E.
Lukens, Nature (London) \textbf{406}, 43 (2000); C. H. van der Wal, A. C. J.
ter Haar, F. K. Wilhelm, R. N. Schouten, C. J. P. M. Harmans, T. P. Orlando,
S. Lloyd, and J. E. Mooij, Science \textbf{290}, 773 (2000); A. J. Berkley,
H. Xu, R. C. Ramos, M. A. Gubrud, F. W. Strauch, P. R. Johnson, J. R.
Anderson, A. J. Dragt, C. J. Lobb, and F. C. Wellstood, Science \textbf{300}%
, 1548 (2003); H. Xu, F. W. Strauch, S. K. Dutta, P. R. Johnson, R. C.
Ramos, A. J. Berkley, H. Paik, J. R. Anderson, A. J. Dragt, C. J. Lobb, and
F. C. Wellstood, Phys. Rev. Lett. \textbf{94}, 027003 (2005).

\bibitem{Rabi} D. Vion, A. Aassime, A. Cottet, P. Joyez, H. Pothier, C.
Urbina, D. Esteve, and M. H. Devoret, Science \textbf{296}, 886 (2002);
Chiorescu, Y. Nakamura, C. J. P. M. Harmans, and J. E. Mooij, Science
\textbf{299}, 1869 (2003).

\bibitem{NHR} A. N. Cleland and M. R. Geller, Phys. Rev. Lett. \textbf{93},
070501 (2004).

\bibitem{TL} A. Wallraff, D. I. Schuster, A. Blais, L. Frunzio, R. S. Huang,
J. Majer, S. Kumar, S. M. Girvin, and R. J. Schoelkopf, Nature (London)
\textbf{431}, 162 (2004).

\bibitem{EITinSQC1} K. V. R. M. Murali, Z. Dutton, W. D. Oliver, D. S.
Crankshaw, and T. P. Orlando, Phys. Rev. Lett. \textbf{93}, 087003 (2004);
Z. Dutton, K. V. R. M. Murali, W. D. Oliver, and T. P. Orlando, Phys. Rev. B
\textbf{73}, 104516 (2006).

\bibitem{EITinSQC2} A. A. Abdumalikov Jr., O. Astafiev, A. M. Zagoskin, Yu. A.
Pashkin, Y. Nakamura, and J. S. Tsai, Phys. Rev. Lett. 104, 193601
(2010).

\bibitem{ATinSQC} M. Baur, S. Filipp, R. Bianchetti, J. M. Fink, M. Goppl,
L. Steffen, P. J. Leek, A. Blais, and A. Wallraff, Phys. Rev. Lett. \textbf{%
102}, 243602 (2009); M. A. Sillanpa, Jian Li, K. Cicak, F. Altomare, J. I.
Park, R. W. Simmonds, G. S. Paraoanu, and P. J. Hakonen, Phys. Rev. Lett.
\textbf{103}, 193601 (2009).

\bibitem{CPTinSQC} W. R. Kelly, Z. Dutton, J. Schlafer, B. Mookerji, T. A.
Ohki, J. S. Kline and D. P. Pappas, Phys. Rev. Lett. \textbf{104}, 163601
(2010).

\bibitem{TsaiScience2010} O. Astafiev, A. M. Zagoskin, A. A. Abdumalikov
Jr., Y. A. Pashkin, T. Yamamoto, K. Inomata, Y. Nakamura and J. S. Tsai,
Science \textbf{327}, 840 (2010).
\end{thebibliography}
\end{document}